\documentclass[epsfig,12pt]{article}
\usepackage{amsfonts}
\usepackage{amssymb}
\usepackage{graphicx}
\usepackage{amsmath}

\input epsf.sty
\newtheorem{theorem}{Theorem}
\newtheorem{acknowledgement}[theorem]{Acknowledgement}

\input{tcilatex}
\makeatletter \@addtoreset{equation}{section}
\renewcommand{\theequation}{\thesection.\arabic{equation}}
\begin{document}

\title{\rightline{\mbox{\small
{Lab/UFR-HEP0803/GNPHE/0803/VACBT/0803}}} \textbf{Refining the Shifted
Topological Vertex}}
\author{L. B Drissi{\small \thanks{%
drissilb@gmail.com}}, H. Jehjouh{\small \thanks{%
jehjouh@gmail.com}}, E.H Saidi{\small \thanks{%
h-saidi@fsr.ac.ma}} \and \vspace{-0.2cm} {\small 1. Lab/UFR- Physique des
Hautes Energies, Facult\'{e} des Sciences, Rabat, Morocco,} \and {\small 2.
GNPHE, Groupement National de Physique des Hautes Energies, Si\`{e}ge focal:
FS, Rabat.}}
\maketitle

\begin{abstract}
We study aspects of the refining and shifting properties of the 3d MacMahon
function $\mathcal{C}_{3}\left( q\right) $ used in topological string theory
and BKP hierarchy. We derive the explicit expressions of the shifted
topological vertex $\mathcal{S}_{\lambda \mu \nu }\left( q\right) $ and its
refined version $\mathcal{T}_{\lambda \mu \nu }\left( q,t\right) $. These
vertices complete results in literature.

\end{abstract}

%\tableofcontents

\section{Introduction}

In the last few years, there has been some interest in the study of the
topological vertex formalism of toric Calabi-Yau threefold (CY3) \textrm{%
\cite{vafa}}. This interest has followed the basic result according to which
the topological vertex $\mathcal{C}_{\lambda \mu \nu }$ is a powerful tool
to compute toric CY3 topological string amplitudes \textrm{\cite%
{vafa,AM,ok,lhh}. }It also came\textrm{\ }from the remarkable relation
between $\mathcal{C}_{\lambda \mu \nu }$ and the Gromov-Witten invariants of
genus g- curves in toric Calabi-Yau threefolds \textrm{\cite{GW,GX,GY}}. 
\newline
Recently it has been shown that the \emph{standard} topological 3- vertex $%
\mathcal{C}_{\lambda \mu \nu }\left( q\right) $ may have two kinds of
generalizations; one known as the refining of $\mathcal{C}_{\lambda \mu \nu
}\left( q\right) $; and the other as its shifting. \newline
In the first case, the \emph{refined} topological vertex $\mathcal{R}%
_{\lambda \mu \nu }\left( q,t\right) $ is a two parameters function
computing the refined topological string amplitudes of toric CY3s \textrm{%
\cite{refined,TA,G}}. It has been found also that $\mathcal{R}_{\lambda \mu
\nu }\left( q,t\right) $ computes as well \ the Nekrasov's instantons of the
topological string free energy $F\left( X^{I},\varepsilon _{1},\varepsilon
_{2}\right) $ of four dimensional $SU\left( N\right) $ gauge theories 
\textrm{\cite{N,N1}}. In Nekrasov's extension, the usual topological string
coupling constant $g_{top}=\ln \left( \frac{1}{q}\right) $ gets replaced by
the pair of parameters $\varepsilon _{1}$, $\varepsilon _{2}$ \textrm{\cite%
{vafa link}}.\newline
In the second case, the \emph{standard} MacMahon function $\mathcal{C}%
_{3}\left( q\right) $ \textrm{\cite{ok,DJS}}\ has been extended to the so
called shifted partition function $\mathcal{S}_{3}\left( q\right) $. This is
the generating function of the shifted plane partitions and it is used in
the study of BKP hierarchy \textrm{\cite{Foda,ST,SS,SSZ}}.

The aim of this paper is to contribute to this matter by combining both the
refining and the shifting operations to get the \emph{refined-shifted}
topological vertex $\mathcal{T}_{\lambda \mu \nu }\left( q,t\right) $
extending $\mathcal{R}_{\lambda \mu \nu }\left( q,t\right) $ and $\mathcal{S}%
_{3}\left( q\right) $ obtained recently in literature. More precisely, we
want to complete the missing relations presented in the two following tables:%
\newline
\textbf{(i)} First, we determine the refined version of the shifted MacMahon
function $\mathcal{S}_{3}\left( q\right) $ obtained by Foda and Wheeler. The
refined version of $\mathcal{S}_{3}\left( q\right) $, denoted below as $%
\mathcal{T}_{3}\left( q,t\right) $, is missing. It is a two parameters
function generating shifted 3d- partitions needed to complete the table,

\begin{equation}
\begin{tabular}{|lll|}
\hline
$\qquad \mathcal{C}_{3}\left( q\right) =\emph{known:}$ \ eq(\ref{m3}) & $%
\quad \underrightarrow{\text{{\small refining}}}$ & $\qquad \mathcal{R}%
_{3}\left( q,t\right) =\emph{known:}$ eq(\ref{r}) \\ 
{\small shifting }$\qquad \downarrow $ &  & $\qquad \qquad \downarrow \qquad 
${\small shifting} \\ 
$\qquad \mathcal{S}_{3}\left( q\right) =\emph{known:}$ \ eq(\ref{sm}) & $%
\quad \underrightarrow{\text{{\small refining}}}$ & $\qquad \mathcal{T}%
_{3}\left( q,t\right) =$ $\ \mathbf{...}\ \mathbf{?}$ \\ \hline
\end{tabular}
\label{1}
\end{equation}

\ \ \ \newline
(\textbf{ii}) Second, we extend the generalized MacMahon functions $\mathcal{%
S}_{3}\left( q\right) $ and $\mathcal{T}_{3}\left( q,t\right) $ by
implementing boundary conditions captured by the strict\footnote{%
Notice that $\xi $ refers to a 2d- partition and $\widehat{\xi }$ to a \emph{%
strict} 2d-partition. The hat is sometimes dropped out for simplicity of
notations.} 2d partitions $\widehat{\lambda },$ $\widehat{\mu }$ and $%
\widehat{\nu }$. The resulting $\mathcal{S}_{\widehat{\lambda }\widehat{\mu }%
\widehat{\nu }}\left( q\right) $ and $\mathcal{T}_{\widehat{\lambda }%
\widehat{\mu }\widehat{\nu }}\left( q,t\right) $ are also needed to complete
the following table,

\begin{equation}
\begin{tabular}{|lll|}
\hline
$\qquad \mathcal{C}_{\lambda \mu \nu }\left( q\right) =\emph{known}$ $\ $eq(%
\ref{v}) & $\quad \underrightarrow{\text{{\small refining}}}$ & $\qquad 
\mathcal{R}_{\lambda \mu \nu }\left( q,t\right) =\emph{known:}$ \ eq(\ref{h1}%
) \\ 
{\small shifting }\ $\ \downarrow $ &  & $\qquad \quad \downarrow \quad $%
{\small shifting} \\ 
$\qquad \mathcal{S}_{\widehat{\lambda }\widehat{\mu }\widehat{\nu }}\left(
q\right) =$ $\ \mathbf{...}\ \mathbf{?}$ & $\quad \underrightarrow{\text{%
{\small refining}}}$ & $\qquad \mathcal{T}_{\widehat{\lambda }\widehat{\mu }%
\widehat{\nu }}\left( q,t\right) =$ $\ \mathbf{...}\ \mathbf{?}$ \\ \hline
\end{tabular}
\label{2}
\end{equation}

\ \newline
The organization of this paper is as follows: In section 2, we give
generalities on topological vertices. In particular, we review briefly the
expression of the constructions of the standard topological vertex $\mathcal{%
C}_{\lambda \mu \nu }\left( q\right) $, the refined one $\mathcal{R}%
_{\lambda \mu \nu }\left( q,t\right) $ and the shifted MacMahon function $%
\mathcal{S}_{3}\left( q\right) $. In section 3, we derive the explicit
expression of the shifted topological vertex $\mathcal{S}_{\widehat{\lambda }%
\widehat{\mu }\widehat{\nu }}\left( q\right) $ of eq(\ref{2}). In section 4,
we compute the its refined version $\mathcal{T}_{\widehat{\lambda }\widehat{%
\mu }\widehat{\nu }}\left( q,t\right) $. In section 5, we give a conclusion
and in section 6, we collect some useful tools as an appendix.

\section{Topological vertex: a review}

In this section, we review briefly some basic tools; in particular the
explicit expressions of the three following topological vertices: \newline
(\textbf{1}) the \emph{standard} topological vertex denoted as $\mathcal{C}%
_{\lambda \mu \nu }\left( q\right) $. \newline
(\textbf{2}) the \emph{refined} topological vertex$\ \mathcal{R}_{\lambda
\mu \nu }\left( q,t\right) $. This is a two parameters generalization of $%
\mathcal{C}_{\lambda \mu \nu }\left( q\right) $. \newline
(\textbf{3}) the \emph{standard} 3d- MacMahon function$\ \mathcal{C}%
_{3}\left( q\right) $, its \emph{refined} version $\mathcal{R}_{3}\left(
q,t\right) $ as well as the\emph{\ shifted} 3d- MacMahon function $\mathcal{S%
}_{3}\left( q\right) $ obtained in \cite{Foda}.

\ \newline
These objects have interpretations in: (\textbf{a}) the topological string
A- model in which $q=e^{-g_{s}}$ with $g_{s}$ being the topological string
coupling constant. (\textbf{b}) the statistical mechanical models in which
the parameter $q$ describes the Boltzmann weight $q=e^{-\frac{1}{KT}}$ with $%
T$ being absolute temperature \textrm{\cite{MS,DJS}}.\newline

\textbf{(a)} \emph{Vertex} $\mathcal{C}_{\lambda \mu \nu }\left( q\right) $%
\newline
Following \textrm{\cite{ok}}, the \emph{standard} topological vertex $%
\mathcal{C}_{\lambda \mu \nu }\left( q\right) $, with boundary conditions in
the $\left( x_{i},x_{j}\right) $ orthogonal planes of $\mathbb{Z}^{3}$
lattice given by the 2d partitions $\left( \lambda ,\mu ,\nu \right) $, can
be defined in the transfer matrix method as follows: 
\begin{equation}
\mathcal{C}_{\lambda \mu \nu }(q)=\mathrm{f}_{\lambda \mu \nu }\left\langle
\nu ^{t}\right\vert \left( \prod\limits_{t=0}^{\infty }q^{L_{0}}\Gamma
_{+}(q^{-\lambda })\right) q^{L_{0}}\left( \prod\limits_{t=-\infty
}^{-1}\Gamma _{-}(q^{-\lambda ^{t}})q^{L_{0}}\right) \left\vert \mu
\right\rangle ,  \label{v}
\end{equation}%
where $\left\vert \xi \right\rangle $ is a generic 2d partition state ( a
Young diagram), $\xi ^{t}$ its transpose and $\left\langle \xi \right\vert
L_{0}\left\vert \xi \right\rangle =\left\vert \xi \right\vert $ with $%
\left\vert \xi \right\vert $ being the number of boxes of the Young diagram.
The operators $\Gamma _{\pm }\left( x\right) $ are vertex operators of the $%
c=1$ bosonic CFT$_{2}$, whose explicit expressions can be found in $\mathrm{%
\cite{DJS}}$, and $\mathrm{f}_{\lambda \mu \nu }$ is a real number given by 
\begin{equation}
\mathrm{f}_{\lambda \mu \nu }=\frac{q^{\frac{1}{2}(\left\Vert \lambda
\right\Vert ^{2}+\left\Vert \mu \right\Vert ^{2}+\left\Vert \nu \right\Vert
^{2})}}{q^{n(\nu )+n(\mu )}}\left( \prod\limits_{n=0}^{\infty }\left(
1-q^{n}\right) ^{n}\right) .
\end{equation}%
For the particular case where there is no boundary condition, i.e $\lambda
=\mu =\nu =\emptyset $, the vertex $\mathcal{C}_{\emptyset \emptyset
\emptyset }\left( q\right) $ coincides exactly with the $3d$- MacMahon
function $\mathcal{C}_{3}$: 
\begin{equation}
\mathcal{C}_{3}\left( q\right) =\left\langle \emptyset \right\vert \left(
\prod\limits_{t=0}^{\infty }q^{L_{0}}\Gamma _{+}(1)\right) q^{L_{0}}\left(
\prod\limits_{t=-\infty }^{-1}\Gamma _{-}(1)q^{L_{0}}\right) \left\vert
\emptyset \right\rangle .
\end{equation}%
The function $\mathcal{C}_{3}$, which reads explicitly as; see also eq(\ref%
{1}), 
\begin{equation}
\mathcal{C}_{3}\left( q\right) =\prod_{n=1}^{\infty }\left( \frac{1}{1-q^{n}}%
\right) ^{n},  \label{m3}
\end{equation}%
has several interpretations. It is the amplitude of the A-model topological 
\emph{closed string} on $\mathbb{C}^{3};$ i.e $\mathcal{C}_{3}=\mathcal{C}%
_{\emptyset \emptyset \emptyset }$. It is also the generating function of 3d
partitions, 
\begin{equation}
\mathcal{C}_{3}\left( q\right) =\dsum\limits_{\text{3d-partitions }\Pi
}q^{\left\vert \Pi \right\vert }\text{ }.
\end{equation}%
Likewise, the vertex $\mathcal{C}_{\lambda \mu \nu }\left( q\right) $
inherits the interpretations of $\mathcal{C}_{3}\left( q\right) $ with a
slight generalization and more power since it allows gluing \textrm{\cite%
{vafa} }to topological amplitudes of all non compact toric CY3s. \newline
It is the partition function of A- model topological string of $\mathbb{C}%
^{3}$ with \emph{open} \emph{strings} on boundaries and, up on using the
gluing method \textrm{\cite{vafa}}, it allows to compute the partition
function of toric CY3s. \newline
$\mathcal{C}_{\lambda \mu \nu }\left( q\right) $ has also a combinatorial
interpretation in terms of the generating function of the plane partitions
with the boundary conditions $\left( \lambda ,\mu ,\nu \right) $ where $%
\lambda ,$ $\mu $ and $\nu $ are $2d$- partitions. \newline
The explicit expression of $\mathcal{C}_{\lambda \mu \nu }$ can be exhibited
in different, but equivalent forms. Its expression in terms of the product
of three Schur functions can be found in \textrm{\cite{ok,refined}}.\newline

\emph{Extensions}\newline
Two kinds of generalizations of the topological vertex $\mathcal{C}_{\lambda
\mu \nu }$ have been considered in the literature. These are: \newline
(\textbf{i}) the refined vertex $\mathcal{R}_{\lambda \mu \nu }\left(
q,t\right) $ having a connection with Nekrasov's partition\textrm{\ }%
function of $SU\left( N\right) $ gauge theories \textrm{\cite{N}} and with
the link invariants \textrm{\cite{vafa link}. }\newline
\textbf{(ii)}\textrm{\ }the shifted MacMahon function\textrm{\ }$\mathcal{S}%
_{3}\left( q\right) $\textrm{\ }used in BKP hierarchy \textrm{\cite{Foda}}.
The $\mathcal{S}_{\lambda \mu \nu }\left( q\right) $ extension of the
shifted vertex by implementing boundary conditions was not computed before;
it is a result of the present paper.\newline
Let us give some details on $\mathcal{R}_{\lambda \mu \nu }$ and $\mathcal{S}%
_{3}$; then we turn back to the computation of $\mathcal{S}_{\lambda \mu \nu
}\left( q\right) $.\newline

\textbf{(b)} \emph{Refined} \emph{vertex:} $\mathcal{R}_{\lambda \mu \nu
}\left( q,t\right) $\newline
The refined topological vertex $\mathcal{R}_{\lambda \mu \nu }\left(
q,t\right) $ is a two parameter extension of $\mathcal{C}_{\lambda \mu \nu
}\left( q\right) $. As noted before, it has a topological string
interpretation in connection with Nekrasov's instantons. Its explicit
expression has been first derived by Iqbal, Kozcaz and Vafa and can be
expressed in different; but equivalent ways. It is given, in the transfer
matrix method, by 
\begin{equation}
\mathcal{R}_{\lambda \mu \nu }\left( q,t\right) =\mathrm{r}_{\lambda \mu \nu
}\left\langle \nu ^{t}\right\vert \left( \prod\limits_{t=0}^{\infty
}t^{L_{0}}\Gamma _{+}(q^{-\lambda _{i}})\right) t^{L_{0}}\left(
\prod\limits_{t=-\infty }^{-1}\Gamma _{-}(q^{-\lambda
_{j}^{t}})q^{L_{0}}\right) \left\vert \mu \right\rangle ,  \label{h1}
\end{equation}%
where $\left\vert \xi \right\rangle $, $q^{L_{0}}$ and $\Gamma _{\pm }\left(
x\right) $ are as before; and 
\begin{equation}
\mathrm{r}_{\lambda \mu \nu }=\frac{q^{\frac{1}{2}(\left\Vert \lambda
\right\Vert ^{2}+\left\Vert \mu \right\Vert ^{2}+\left\Vert \nu \right\Vert
^{2})}}{q^{n(\nu ^{t})}t^{n(\mu )}}\prod_{k,l=1}^{\infty }(1-q^{k-1}t^{l}).
\end{equation}%
For the particular case $\lambda =\mu =\nu =\emptyset $, the vertex $%
\mathcal{R}_{\emptyset \emptyset \emptyset }\left( q\right) $ coincides
exactly with the refined $3d$- MacMahon function mentioned in the
introduction (\ref{1}), 
\begin{equation}
\mathcal{R}_{3}\left( q,t\right) =\prod_{k,l=1}^{\infty }(1-q^{k-1}t^{l}).
\label{r}
\end{equation}%
By setting $t=q$ back in (\ref{r}), we get the standard $\mathcal{C}%
_{3}\left( q\right) $ relation. The explicit expression of the refined $%
\mathcal{R}_{\lambda \mu \nu }\left( q,t\right) $ in terms of Schur
functions can be found in \textrm{\cite{refined}}.\newline

(\textbf{3}) \emph{Shifted} $\mathcal{S}_{3}\left( q\right) $ \newline
The shifted 3d MacMahon $\mathcal{S}_{3}(q)$ is the generating functional of 
\emph{strict }plane partitions. The explicit expression of $\mathcal{S}%
_{3}(q $ has been derived by Foda and Wheeler by using transfer matrix
method. It reads as, 
\begin{equation}
\mathcal{S}_{3}(q)=\dprod\limits_{n=1}^{\infty }\left( \frac{1+q^{n}}{1-q^{n}%
}\right) ^{n}\text{ },  \label{sm}
\end{equation}%
and has an interpretation in the BKP hierarchy of the so called neutral free
fermions. It was claimed in \textrm{\cite{Foda}} that $\mathcal{S}_{3}(q)$
could be relevant to the topological string dual to $O(N)$ Chern-Simon
theory in the limit $N\rightarrow \infty $.

\section{Shifted topological vertex}

The expression eq(\ref{sm}) of the \emph{shifted }topological vertex has
been derived in the absence of any kind of boundary conditions. Here, we
want to complete this result by considering the derivation of the \emph{%
shifted }topological vertex $\mathcal{S}_{\lambda \mu \nu }$ with generic
boundary conditions with the property, 
\begin{equation}
\mathcal{S}_{\lambda \mu \nu }=\mathcal{S}_{\lambda \mu \nu }\left( q\right) 
\text{ \qquad ,\qquad }\mathcal{S}_{\emptyset \emptyset \emptyset }=\mathcal{%
S}_{3}\left( q\right) \text{ }.
\end{equation}%
Notice that $\mathcal{S}_{\lambda \mu \nu }\left( q\right) $ generates the
shifted 3d partitions with boundary conditions given by the \emph{strict} 2d
partitions $\left( \lambda ,\mu ,\nu \right) $ along the axis $\left(
x_{1},x_{2},x_{3}\right) $. For the definitions of the shifted 3d- and
strict 2d partitions; see appendix.\newline
The main result of this section is collected in the following proposition
where some terminology has been borrowed from \textrm{\cite{refined}}:%
\newline

\emph{Proposition \textbf{1} }\newline
The perpendicular shifted topological vertex $\mathcal{S}_{\lambda \mu \nu
}\left( q\right) $ with generic boundary conditions, given by three strict
2d- partitions $\left( \lambda ,\mu ,\nu \right) $, reads as follows: 
\begin{equation}
\mathcal{L}_{\lambda \mu \nu }\left( q\right) =\mathrm{f}_{\lambda \mu \nu
}\times \mathcal{S}_{\lambda \mu \nu }\times \left(
\dprod\limits_{n=1}^{\infty }\left( \frac{1-q^{n}}{1+q^{n}}\right)
^{n}\right) \text{ },  \label{sh}
\end{equation}%
which can be also put in the normalized form 
\begin{equation}
\begin{tabular}{llll}
$\mathcal{S}_{\lambda \mu \nu }$ & $=$ & $\mathcal{S}_{\emptyset \emptyset
\emptyset }\mathcal{L}_{\lambda \mu \nu }$ & , \\ 
$\mathcal{L}_{\emptyset \emptyset \emptyset }\left( q\right) $ & $=$ & $1$ & 
.%
\end{tabular}%
\end{equation}
In the relation (\ref{sh}), the numerical factor $\mathrm{f}_{\lambda \mu
\nu }$ is, 
\begin{equation}
\begin{tabular}{llll}
$\mathrm{f}_{\lambda \mu \nu }$ & $=$ & $2^{l(\mu )+l(\lambda )+l(\nu )}q^{%
\frac{1}{2}(\left\Vert \lambda \right\Vert ^{2}+\left\Vert \mu \right\Vert
^{2}+\left\Vert \nu \right\Vert ^{2})}$ & , \\ 
$\mathrm{f}_{\emptyset \emptyset \emptyset }$ & $=$ & $1$ & ,%
\end{tabular}
\label{h2}
\end{equation}%
where $\left\Vert \mu \right\Vert ^{2}=\sum_{i}\mu _{i}^{2}$ and $l(\mu )$
is the length of the \emph{strict} partition that is the number parts of the
strict 2d partition $\mu =\left( \mu _{1},...,\mu _{l(\mu )},0,...\right) $.%
\newline
The function $\mathcal{S}_{\lambda \mu \nu }\left( q\right) $ is the
perpendicular partition function of shifted 3d- partitions. It reads in
terms of Schur functions $P_{\lambda ^{t}/\eta }$ and $Q_{\nu /\eta }$ as
follows: 
\begin{equation}
\begin{tabular}{ll}
$\mathcal{S}_{\lambda \mu \nu }$ & $=\left( \dsum\limits_{\text{strict 2d }%
\eta \text{ }}P_{\lambda ^{t}/\eta }\left( q^{-\nu -\rho }\right) Q_{\nu
/\eta }\left( q^{-\nu ^{t}-\rho }\right) \right) \times \mathrm{Z}(\nu
)\times \mathrm{h}_{\lambda \mu }$ \ ,%
\end{tabular}
\label{h3}
\end{equation}%
with%
\begin{equation}
\begin{tabular}{llll}
$\rho $ & $=$ & $\left( \rho _{1},...,\rho _{i},...\right) $ & , \\ 
$\rho _{k}$ & $=$ & $\frac{1}{2}-k$ & , \\ 
$n(\xi )$ & $=$ & $\frac{1}{2}\left( \left\Vert \xi ^{t}\right\Vert
^{2}-\left\vert \xi \right\vert \right) $ & ,%
\end{tabular}%
\end{equation}%
as well as 
\begin{equation}
\begin{tabular}{llll}
$\mathrm{h}_{\lambda \mu }$ & $=$ & $2^{-l(\mu )-l(\lambda )}q^{-n(\lambda
^{t})-n(\mu )}q^{-\frac{\left\vert \mu \right\vert }{2}-\frac{\left\vert
\lambda \right\vert }{2}}$ & , \\ 
$q^{-\nu -\rho }$ & $=$ & $\left( q^{-\nu _{1}-\rho _{1}},...,q^{-\nu
_{i}-\rho _{i}},...\right) $ & ,%
\end{tabular}
\label{h4}
\end{equation}%
and 
\begin{equation}
\begin{tabular}{lll}
$\mathrm{Z}(\nu )$ & $=\frac{2^{-l(\nu )}P_{\nu ^{t}}(q^{-\rho })}{%
q^{+\left( \left\vert \nu \right\vert /2\right) +n(\nu ^{t})}}%
\dprod\limits_{n=1}^{\infty }\left( \frac{1+q^{n}}{1-q^{n}}\right) ^{n}$ & $%
. $%
\end{tabular}%
\end{equation}%
The function $P_{\xi }\left( x\right) $ is the Schur function associated
with the strict 2d partition $\xi $; it is defined as 
\begin{equation}
\Gamma _{-}\left( x\right) \left\vert \lambda \right\rangle =\dsum\limits_{%
\text{strict 2d partition }{\small \xi >\lambda }}P_{\xi /\lambda }\left(
x\right) \left\vert \xi \right\rangle .  \label{38}
\end{equation}%
where $\xi /\lambda $ is the complement of $\lambda $ in $\xi $. We also
have the orthogonality relation 
\begin{equation}
\begin{tabular}{llll}
$Q_{\xi }\left( x\right) $ & $=$ & $2^{-l\left( \xi \right) }P_{\xi }\left(
x\right) $ & , \\ 
$\left\langle Q_{\xi }\left( x\right) ,P_{\zeta }\left( x\right)
\right\rangle $ & $=$ & $\delta _{\xi \zeta }$ & .%
\end{tabular}
\label{39}
\end{equation}%
where $\delta _{\xi \zeta }=\tprod_{i}\delta _{\xi _{i}\zeta _{i}}$.

\ \ \ \newline
To establish this result, we consider shifted 3d- partitions $\Pi ^{\left(
3\right) }$ inside of a cube with size $N_{1}N_{2}N_{3}$ and boundary
conditions given by the strict 2d- partitions $\left( \lambda ,\mu ,\nu
\right) $. More precisely, the strict 2d- partition $\lambda $ belongs to
the plane $\left( x_{2},x_{3}\right) $ of the ambient real 3-dimensional
space, $\mu $ belongs to the plane $\left( x_{3},x_{1}\right) $ and $\nu $
to the plane $\left( x_{1},x_{2}\right) $, \newline
Then proceed by steps as follows:\newline

\emph{Step one}: \qquad \newline
Compute the perpendicular partition function $\mathcal{S}_{\lambda \mu \nu
}\left( q\right) $\ by using the transfer matrix approach \textrm{\cite{ok}}%
. This method has been used for calculating the topological vertex $\mathcal{%
C}_{\lambda \mu \nu }$ of the A- model topological string on $\mathbb{C}^{3}$
which lead to eqs(\ref{v}-\ref{m3}). $\mathcal{S}_{\lambda \mu \nu }$ reads
in terms of products $\Gamma _{\pm }\left( x\right) $ as follows 
\begin{equation}
\mathcal{S}_{\lambda \mu \nu }=\frac{q^{-n\left( \nu \right) -n\left( \mu
^{t}\right) }}{q^{N_{2}\left\vert \mu \right\vert +N_{1}\left\vert \nu
\right\vert }}\left\langle \nu ^{t}|\left(
\prod\limits_{j=1}^{N_{1}}q^{L_{0}}\Gamma _{+}\left( q^{-\lambda
_{j}}\right) \right) q^{L_{0}}\left( \prod\limits_{i=1}^{N_{2}}\Gamma
_{-}\left( q^{-\lambda _{i}^{t}}\right) q^{L_{0}}\right) |\mu \right\rangle 
\text{ }.
\end{equation}%
By using the relation $q^{-kL_{0}}\Gamma _{\pm }\left( z\right)
q^{kL_{0}}=\Gamma _{\pm }\left( zq^{k}\right) $ and $q^{L_{0}}\left\vert \xi
\right\rangle =2^{l\left( \xi \right) }q^{\left\vert \xi \right\vert
}\left\vert \xi \right\rangle $, the function $\mathcal{S}_{\lambda \mu \nu
} $ can be brought to 
\begin{equation}
\mathcal{S}_{\lambda \mu \nu }=\vartheta _{\mu \nu }\left\langle \nu
^{t}|\left( \prod\limits_{j=1}^{N_{1}}\Gamma _{+}\left( q^{-\lambda
_{j}-\rho _{j}}\right) \right) \left( \prod\limits_{i=1}^{N_{2}}\Gamma
_{-}\left( q^{-\lambda _{i}^{t}-\rho _{i}}\right) \right) |\mu \right\rangle 
\text{ },
\end{equation}%
with 
\begin{equation}
\vartheta _{\mu \nu }=\frac{q^{-\left( \frac{\left\vert \nu \right\vert }{2}%
+n\left( \nu \right) \right) -\left( \frac{\left\vert \mu \right\vert }{2}%
+n\left( \mu ^{t}\right) \right) }}{2^{l\left( \nu \right) +l\left( \mu
\right) }}\text{ }.
\end{equation}%
The vertex operators $\Gamma _{\pm }(z)$ are given by 
\begin{equation}
\begin{tabular}{llll}
$\Gamma _{+}(z)$ & $=$ & $\exp \left( \sum_{m\in \mathbb{N}_{odd}}\frac{2}{m}%
z^{m}\lambda _{m}\right) $ & , \\ 
$\Gamma _{-}(z)$ & $=$ & $\exp \left( \sum_{m\in \mathbb{N}_{odd}}\frac{2}{m}%
z^{m}\lambda _{-m}\right) $ & ,%
\end{tabular}
\label{ga}
\end{equation}%
with $\lambda _{m}$ being operators satisfying the following commutation
relations, 
\begin{equation}
\left[ \lambda _{m},\lambda _{n}\right] =-\frac{m}{2}\delta _{n+m,0}\qquad
,\qquad m,\text{ }n\in \mathbb{Z}_{odd}\text{ }.
\end{equation}%
Notice that in the particular limit $N_{1},$ $N_{2}$ and $N_{3}\rightarrow
\infty $, the vertex $\mathcal{S}_{\emptyset \emptyset \emptyset }$ gets
identified with 
\begin{equation}
\mathcal{S}_{3}\left( q\right) =\sum_{\text{shifted 3d }\Pi }2^{p(\Pi
)}q^{\left\vert \Pi \right\vert }\text{ ,}  \label{foda}
\end{equation}%
whose explicit expression is precisely (\ref{sm}).\newline

\emph{Step two}:\qquad \newline
Commuting the vertex operators $\Gamma _{+}\left( z_{i}\right) $ to the
right of $\Gamma _{-}\left( z_{j}\right) $ by using the commutation
relations 
\begin{equation}
\begin{tabular}{llll}
$\Gamma _{+}(x)\Gamma _{-}(y)$ & $=$ & $\left( \frac{1+xy}{1-xy}\right)
\Gamma _{-}(y)\Gamma _{+}(x)$ & , \\ 
$\Gamma _{\pm }(x)\Gamma _{\pm }(y)$ & $=$ & $\Gamma _{\pm }(y)\Gamma _{\pm
}(x)$ & ,%
\end{tabular}
\label{RQ}
\end{equation}%
we get 
\begin{equation}
\mathcal{S}_{\lambda \mu \nu }=\zeta _{\lambda \mu \nu }\dsum\limits_{\text{%
strict 2d }\eta \text{ }}\left\langle \nu ^{t}|\left(
\prod\limits_{j=1}^{N_{1}}\Gamma _{-}\left( q^{-\lambda _{j}^{t}-\rho
_{j}}\right) \right) \left\vert \eta \right\rangle \left\langle \eta
\right\vert \left( \prod\limits_{i=1}^{N_{2}}\Gamma _{+}\left( q^{-\lambda
_{i}-\rho _{i}}\right) \right) |\mu \right\rangle \text{ },
\end{equation}%
with 
\begin{equation}
\zeta _{\lambda \mu \nu }=\mathrm{Z}(\lambda )\frac{q^{-\left( \frac{%
\left\vert \nu \right\vert }{2}+n\left( \nu \right) \right) -\left( \frac{%
\left\vert \mu \right\vert }{2}+n\left( \mu ^{t}\right) \right) }}{%
2^{l\left( \nu \right) +l\left( \mu \right) }}\text{ }.
\end{equation}%
This relation can be simplified further by using the Schur functions $P_{\nu
^{t}/\eta }\left( x_{j}\right) =\left\langle \nu ^{t}\right\vert \Gamma
_{-}(x_{j})\left\vert \eta \right\rangle $ and $Q_{\nu ^{t}/\eta
}(x_{i})=\left\langle \nu ^{t}\right\vert \Gamma _{+}(x_{j})\left\vert \eta
\right\rangle $ following from the identities $\left\langle \nu
^{t}\right\vert \tprod_{j=1}^{N_{2}}\Gamma _{-}(x_{j})\left\vert \eta
\right\rangle =P_{\lambda /\eta }(x_{1},...,x_{N})$, $\left\langle \nu
^{t}\right\vert \tprod_{j=1}^{N_{1}}\Gamma _{=}(x_{j})\left\vert \eta
\right\rangle =Q_{\lambda /\eta }(x_{1},...,x_{N})$ as well as, 
\begin{equation}
\begin{tabular}{llll}
$\Gamma _{\pm }(x_{j})\left\vert \eta \right\rangle $ & $=$ & $\dsum\limits_{%
\text{strict 2d }\lambda >\eta \text{ }}L_{\lambda /\eta }^{\pm
}(x_{j})\left\vert \lambda \right\rangle $ & , \\ 
$\dprod\limits_{j=1}^{N_{1}}\Gamma _{\pm }(x_{j})\left\vert \eta
\right\rangle $ & $=$ & $\dsum\limits_{\text{strict 2d }\lambda >\eta \text{ 
}}L_{\lambda /\eta }^{\pm }(x_{1},x_{2},\ldots )\left\vert \lambda
\right\rangle $ & ,%
\end{tabular}
\label{dd}
\end{equation}%
with $x=\left( x_{1},x_{2},...\right) $ and where we have set $L_{\lambda
/\eta }^{-}=P_{\lambda /\eta }$ and $L_{\lambda /\eta }^{+}=Q_{\lambda /\eta
}$.\newline
Then, the partition function $\mathcal{S}_{\lambda \mu \nu }$ becomes, 
\begin{equation}
\mathcal{S}_{\lambda \mu \nu }\left( q\right) =Z(\lambda )\frac{q^{-\left( 
\frac{\left\vert \nu \right\vert }{2}+n\left( \nu \right) \right) -\left( 
\frac{\left\vert \mu \right\vert }{2}+n\left( \mu ^{t}\right) \right) }}{%
2^{l\left( \nu \right) +l\left( \mu \right) }}\dsum\limits_{\text{strict 2d }%
\eta \text{ }}P_{\nu ^{t}/\eta }\left( q^{-\lambda ^{t}-\rho }\right) Q_{\mu
/\eta }\left( q^{-\lambda -\rho }\right) .  \label{zl}
\end{equation}%
To determine the factor $\mathrm{Z}(\lambda )$, we first use the identity $%
\mathcal{S}_{\lambda \emptyset \emptyset }=\mathrm{Z}(\lambda )$, then the
cyclic property $\mathcal{S}_{\lambda \mu \nu }=\mathcal{S}_{\mu \nu \lambda
}=\mathcal{S}_{\nu \lambda \mu }$ which implies in turns that $\mathcal{S}%
_{\lambda \emptyset \emptyset }=\mathcal{S}_{\emptyset \emptyset \lambda }$;
from which we learn the following result 
\begin{equation}
\mathrm{Z}(\lambda )=\frac{q^{-\frac{\left\vert \lambda \right\vert }{2}%
-n(\lambda ^{t})}}{2^{l(\lambda )}}\left[ \dprod\limits_{n=1}^{\infty
}\left( \frac{1+q^{n}}{1-q^{n}}\right) ^{n}\right] P_{\lambda ^{t}}(q^{-\rho
})\text{ }.  \label{h6}
\end{equation}

\emph{Step three}:\qquad \newline
The shifted MacMahon function $\mathcal{S}_{3}\left( q\right) $ can be
recovered from the above analysis by using the identity $\mathcal{S}%
_{3}\left( q\right) =\mathcal{S}_{\emptyset \emptyset \emptyset }=\mathrm{Z}%
(\emptyset )$. \newline
This ends the proof of eq(\ref{sh}). Notice that $\mathcal{L}_{\lambda \mu
\nu }\left( q\right) $ can be also put in the form 
\begin{equation}
\mathcal{L}_{\lambda \mu \nu }\left( q\right) =q^{\frac{k(\nu )}{2}%
}P_{\lambda ^{t}}(q^{-\rho })\sum_{\text{ strict 2d }\eta }P_{\lambda
^{t}/\eta }\left( q^{-\nu -\rho }\right) Q_{\nu /\eta }\left( q^{-\nu
^{t}-\rho }\right) \text{ ,}  \label{sr}
\end{equation}%
where $k(\xi )=2(\left\Vert \xi \right\Vert ^{2}-\left\Vert \xi
^{t}\right\Vert ^{2})$ is the Casimir associated to $\xi $ strict 2d
partition.

\section{Refining the shifted vertex}

In this section, we derive the explicit expression of the refining version $%
\mathcal{T}_{\lambda \mu \nu }\left( q,t\right) $ of the shifted topological
vertex $\mathcal{S}_{\lambda \mu \nu }\left( q\right) $. This is a two
parameters $q$ and $t$ with boundary conditions given by the strict 2d
partitions $\lambda ,\mu $ and $\nu $. \newline
Notice that like for $\mathcal{R}_{\lambda \mu \nu }$ of eq(\ref{h1}), the
function the \emph{refined-shifted} topological vertex $\mathcal{T}_{\lambda
\mu \nu }$ is non cyclic with respect to the permutations of the strict 2d
partitions $\left( \lambda ,\mu ,\nu \right) $; $\mathcal{T}_{\lambda \mu
\nu }\neq \mathcal{T}_{\mu \nu \lambda }\neq \mathcal{T}_{\nu \lambda \mu }$%
. It obeys however the properties 
\begin{equation}
\begin{tabular}{llllllll}
$\mathcal{T}_{\lambda \mu \nu }\left( q,q\right) $ & $=$ & $\mathcal{S}%
_{\lambda \mu \nu }\left( q\right) $ & , & $\mathcal{T}_{\emptyset \emptyset
\emptyset }\left( q,t\right) $ & $=$ & $\mathcal{T}_{3}\left( q,t\right) $ & 
.%
\end{tabular}
\label{41}
\end{equation}

\emph{Proposition \textbf{2}:\newline
}The explicit expression of the refined-shifted topological vertex $\mathcal{%
T}_{\lambda \mu \nu }$ reads as follows\emph{\ } 
\begin{equation}
\mathcal{K}_{\lambda \mu \nu }\left( t,q\right) =\mathrm{f}_{\lambda \mu \nu
}\times \mathcal{T}_{\lambda \mu \nu }\times \left[ \prod\limits_{i,j=1}^{%
\infty }\left( \frac{1+q^{j-1}t^{i}}{1-q^{j-1}t^{i}}\right) \right] ,
\label{rsh}
\end{equation}%
where the factor $\mathrm{f}_{\lambda \mu \nu }=\mathrm{f}_{\lambda \mu \nu
}\left( q\right) $ is the same as in eq(\ref{h2}).\newline
$\mathcal{T}_{\lambda \mu \nu }=\mathcal{T}_{\lambda \mu \nu }\left(
q,t\right) $ is the refining version of $\mathcal{S}_{\lambda \mu \nu
}\left( q\right) $ of eq(\ref{h3}); it is the perpendicular partition
function generating the strict plane partitions. Its explicit expression
reads in terms of the skew Schur functions $P_{\lambda ^{t}/\eta }$ and $%
Q_{\mu /\eta }$ as follows, 
\begin{equation}
\mathcal{T}_{\lambda \mu \nu }=\widetilde{\mathrm{h}}_{\lambda \mu
}(q,t)\times Z_{\nu }(t,q)\sum_{\text{stric 2d partitions }\eta }\varepsilon
_{\nu }(q,t)P_{\lambda ^{t}/\eta }\left( q^{-\nu }t^{-\rho }\right) Q_{\mu
/\eta }\left( t^{-\nu ^{t}}q^{-\rho }\right) \text{ .}
\end{equation}%
where $\widetilde{\mathrm{h}}_{\lambda \mu }$ is the refinement of $\mathrm{h%
}_{\lambda \mu }$ (\ref{h4}) and it is given by 
\begin{equation}
\begin{tabular}{llll}
$\widetilde{\mathrm{h}}_{\lambda \mu }(q,t)$ & $=$ & $2^{-l(\mu )-l(\lambda
)}q^{-n(\lambda ^{t})-\frac{\left\vert \mu \right\vert }{2}}t^{-n(\mu )-%
\frac{\left\vert \lambda \right\vert }{2}}$ & .%
\end{tabular}%
\end{equation}%
We also have 
\begin{equation}
\begin{tabular}{ll}
$Z_{\nu }(t,q)$ & $=\left( \frac{q}{t}\right) ^{\frac{\left\Vert \nu
\right\Vert ^{2}}{2}}2^{-l(\nu )}P_{\nu ^{t}}(t^{-\rho
})\prod\limits_{i,j=1}^{\infty }\left( \frac{1+q^{j-1}t^{i}}{1-q^{j-1}t^{i}}%
\right) $ $.$%
\end{tabular}%
\end{equation}%
as well as $\mathrm{\varepsilon }_{\nu }(q,t)=\left( \frac{q}{t}\right) ^{%
\frac{\left\vert \eta \right\vert }{2}}$. Notice that for $q=t=1$, $\mathrm{%
\varepsilon }_{\nu }(q,t)=1.$ \newline
To establish this result, we use the following steps.\newline

\emph{Step one}: \newline
Compute the refined expression $\mathcal{T}_{3}\left( q,t\right) $ of the
shifted $3d$ MacMahon's function $\mathcal{S}_{3}\left( q\right) $ in terms
of the two parameters q and t. To that purpose, we start from the defining
relation of $\mathcal{T}_{3}\left( q,t\right) $ by using strict 2d-
partitions, 
\begin{equation}
\mathcal{T}_{3}\left( q,t\right) =\sum_{\text{strict 2d partition }\pi
}2^{p(\pi )}q^{\left( \sum_{a=1}^{\infty }\left\vert \pi (-a)\right\vert
\right) }t^{\left( \sum_{a=1}^{\infty }\left\vert \pi (a-1)\right\vert
\right) },
\end{equation}%
where we have used the diagonal slicing of shifted 3d- partitions $\Pi $ in
terms of the strict 2d- ones $\pi \left( a\right) $ as shown below 
\begin{equation}
\Pi =\sum_{a\in \mathbb{Z}}\pi \left( a\right) \qquad ,\qquad \pi \left(
a\right) =\sum_{i}\pi _{i,i+a}\text{ }.
\end{equation}%
Notice that the slices with $a<0$ are weighted by the factor $q^{\left\vert
\pi \left( a\right) \right\vert }$ while the slices with $a\geq 0$ are
weighted by $t^{\left\vert \pi \left( a\right) \right\vert }$.\newline
Then, we use the transfer matrix method which allows to express $\mathcal{T}%
_{3}\left( q,t\right) $ as the amplitude $\mathcal{T}_{\emptyset \emptyset
\emptyset }(q,t)$; that is 
\begin{equation}
\begin{array}{cccc}
\mathcal{T}_{\emptyset \emptyset \emptyset }(q,t) & = & \left\langle
0\right\vert \left( \prod\limits_{a=0}^{\infty }t^{L_{0}}\Gamma
_{+}(1)\right) t^{L_{0}}\left( \prod\limits_{a=-1}^{-\infty }\Gamma
_{-}(1)q^{L_{0}}\right) \left\vert 0\right\rangle & .%
\end{array}%
\end{equation}%
By using $q^{-kL_{0}}\Gamma _{\pm }\left( z\right) q^{kL_{0}}=\Gamma _{\pm
}\left( zq^{k}\right) $, we can also put $\mathcal{T}_{\emptyset \emptyset
\emptyset }$ in the form 
\begin{equation}
\begin{array}{cccc}
\mathcal{T}_{\emptyset \emptyset \emptyset }(q,t) & = & \left\langle
0\right\vert \left( \prod\limits_{i\succ 0}\Gamma _{+}(t^{i})\right) \left(
\prod\limits_{j\succ 0}\Gamma _{-}(q^{j-1})\right) \left\vert 0\right\rangle
& .%
\end{array}%
\end{equation}%
Next commuting the $\Gamma _{-}$'s to the left of the $\Gamma _{+}$'s by
help of the relations (\ref{RQ}), we obtain 
\begin{equation}
\mathcal{T}_{3}(q,t)=\mathcal{T}_{\emptyset \emptyset \emptyset
}(q,t)=\prod\limits_{j=1}^{\infty }\prod\limits_{i=1}^{\infty }\left( \frac{%
1+q^{j-1}t^{i}}{1-q^{j-1}t^{i}}\right) \text{ },  \label{h7}
\end{equation}%
which reduces to $\mathcal{S}_{3}\left( q\right) $ eq(\ref{foda}) by setting 
$t=q$.\newline

\emph{Step two}: \newline
To get the expression of the perpendicular partition function $\mathcal{T}%
_{\lambda \mu \nu }$ for arbitrary boundary conditions, we mimic the
approach of \textrm{\cite{vafa}} and factorize $\mathcal{T}_{\lambda \mu \nu
}$ as follows 
\begin{equation}
\mathcal{T}_{\lambda \mu \nu }(q,t)=g_{\lambda \mu }(t,q)\times \mathcal{T}%
_{\lambda \mu \nu }^{diag},  \label{pf}
\end{equation}%
where $\mathcal{T}_{\lambda \mu \nu }^{diag}$ stands for the diagonal
partition function and $g_{\lambda \mu }(t,q)$ given by 
\begin{equation}
g_{\lambda \mu }(t,q)=q^{-n(\lambda ^{t})}t^{-n(\mu )}
\end{equation}%
describing the change from diagonal slicing to perpendicular one. To compute 
$\mathcal{T}_{\lambda \mu \nu }^{\text{{\small diag}}}$, we use the transfer
matrix method. We first have 
\begin{equation}
\begin{array}{cccc}
\mathcal{T}_{\lambda \mu \nu }^{diag} & = & \left\langle \lambda
^{t}\right\vert \left( \prod\limits_{i\succ 0}t^{L_{0}}\Gamma _{+}(q^{-\nu
_{i}})\right) t^{L_{0}}\left( \prod\limits_{j\succ 0}\Gamma _{-}(t^{-\nu
_{j}^{t}})q^{L_{0}}\right) \left\vert \mu \right\rangle & .%
\end{array}%
\end{equation}%
By using $q^{-kL_{0}}\Gamma _{\pm }\left( z\right) q^{kL_{0}}=\Gamma _{\pm
}\left( zq^{k}\right) $, we can bring it to the form, 
\begin{equation}
\begin{array}{cccc}
\mathcal{T}_{\lambda \mu \nu }^{diag} & = & \left\langle \lambda
^{t}\right\vert \left( \prod\limits_{i\succ 0}\Gamma _{+}(t^{i}q^{-\nu
_{i}})\right) \left( \prod\limits_{j\succ 0}\Gamma _{-}(q^{j-1}t^{-\nu
_{j}^{t}})\right) \left\vert \mu \right\rangle & .%
\end{array}%
\end{equation}%
Then using $\left\langle \mu \right\vert q^{L_{0}}\left\vert \mu
\right\rangle =2^{l(\mu )}q^{\left\vert \mu \right\vert }$ and eq(\ref{RQ}),
we end with 
\begin{equation}
\mathcal{T}_{\lambda \mu \nu }^{diag}=\zeta _{\lambda \mu }\times Z_{\nu
}\times \sum_{\eta }\varepsilon _{\nu }\left( \left\langle \lambda
^{t}\right\vert \ \prod\limits_{j\succ 0}\Gamma _{-}(q^{-\nu }t^{-\rho
})\prod\limits_{i\succ 0}\Gamma _{+}(t^{-\nu ^{t}}q^{-\rho })\ \left\vert
\mu \right\rangle \right) .
\end{equation}%
with 
\begin{equation}
\begin{array}{cccc}
\zeta _{\lambda \mu }(q,t) & = & 2^{-l(\mu )-l(\lambda )}q^{-\frac{%
\left\vert \mu \right\vert }{2}}t^{-\frac{\left\vert \lambda \right\vert }{2}%
} & , \\ 
\varepsilon _{\nu }(q,t) & = & \left( \frac{q}{t}\right) ^{\frac{\left\vert
\eta \right\vert }{2}} & .%
\end{array}%
\end{equation}%
Using eqs(\ref{dd}) and the skew Schur functions $P_{\lambda ^{t}/\eta }$
and $Q_{\mu /\eta }$, the partition function (\ref{pf}) reads as, 
\begin{equation}
\mathcal{T}_{\lambda \mu \nu }=\widetilde{\mathrm{h}}_{\lambda \mu }\times
Z_{\nu }\times \sum_{\eta \text{ strict}}\left( \frac{q}{t}\right) ^{\frac{%
\left\vert \eta \right\vert }{2}}\left[ P_{\lambda ^{t}/\eta }\left( q^{-\nu
}t^{-\rho }\right) Q_{\mu /\eta }\left( t^{-\nu ^{t}}q^{-\rho }\right) %
\right] \text{ ,}
\end{equation}%
where 
\begin{equation}
\begin{tabular}{llll}
$\widetilde{\mathrm{h}}_{\lambda \mu }(q,t)$ & $=$ & $\zeta _{\lambda \mu
}(q,t)\times g_{\lambda \mu }(t,q)$ & , \\ 
& $=$ & $2^{-l(\mu )-l(\lambda )}q^{-n(\lambda ^{t})-\frac{\left\vert \mu
\right\vert }{2}}t^{-n(\mu )-\frac{\left\vert \lambda \right\vert }{2}}$ & .%
\end{tabular}%
\end{equation}%
To determine the factor $Z(\nu )$, we need two data: first we use the
identity $\mathcal{T}_{\emptyset \emptyset \nu }=Z_{\nu }(q,t)$ and second,
we require that $Z_{\nu =\emptyset }=\mathcal{T}_{3}(q,t)$, as in eq(\ref{41}%
). We find 
\begin{equation}
Z_{\nu }(q,t)=\left( \frac{q}{t}\right) ^{\frac{\left\Vert \nu \right\Vert
^{2}}{2}}2^{-l(\nu )}P_{\nu ^{t}}(t^{-\rho })\prod\limits_{j,i=1}^{\infty
}\left( \frac{1+q^{j-1}t^{i}}{1-q^{j-1}t^{i}}\right) .
\end{equation}%
This ends the proof of eq(\ref{rsh}). \newline
Notice that $\mathcal{K}_{\lambda \mu \nu }\left( q,t\right) $ can be also
put in the closed form 
\begin{equation}
\mathcal{K}_{\lambda \mu \nu }=\left( \frac{q}{t}\right) ^{\frac{\left\Vert
\mu \right\Vert ^{2}+\left\Vert \nu \right\Vert ^{2}}{2}}t^{\frac{k(\mu )}{2}%
}P_{\nu ^{t}}(t^{-\rho })\sum_{\eta }\left[ \left( \frac{q}{t}\right) ^{%
\frac{\left\vert \eta \right\vert +\left\vert \lambda \right\vert
-\left\vert \mu \right\vert }{2}}P_{\lambda ^{t}/\eta }\left( x_{\nu ,\rho
}\right) Q_{\mu /\eta }\left( y_{\nu ,\rho }\right) \right] ,
\end{equation}%
with $x_{\nu ,\rho }=q^{-\nu }t^{-\rho }$, $y_{\nu ,\rho }=t^{-\nu
^{t}}q^{-\rho }$ and the property $\mathcal{T}_{\lambda \mu \nu }=\mathcal{T}%
_{\emptyset \emptyset \emptyset }\times \mathcal{K}_{\lambda \mu \nu }$ as
well as the normalization $\mathcal{K}_{\emptyset \emptyset \emptyset }=1$.

\section{Conclusion}

In this paper we have studied the refining and the shifting properties of
the standard topological vertex $\mathcal{C}_{\lambda \mu \nu }$. After
having reviewed some basic properties on: \newline
\textbf{(1}) the standard vertex $\mathcal{C}_{\lambda \mu \nu }\left(
q\right) $ and its refined version $\mathcal{R}_{\lambda \mu \nu }$ used in
the framework ot topological strings,\newline
\textbf{(2)} the shifted MacMahon function $\mathcal{S}_{3}\left( q\right) $
used in BKP hierarchy,\newline
we have completed the missing relations in eqs(\ref{1}) and (\ref{2}). In
particular, we have derived the explicit expressions of: \newline
(\textbf{a}) the shifted topological vertex $\mathcal{S}_{\widehat{\lambda }%
\widehat{\mu }\widehat{\nu }}\left( q\right) $ with boundary conditions
given by generic strict 2d partitions. The shifted MacMahon function $%
\mathcal{S}_{3}\left( q\right) $, given by eq(\ref{sm}) and first obtained
in \textrm{\cite{Foda}}, follows by putting $\widehat{\lambda }=\widehat{\mu 
}=\widehat{\nu }=\emptyset $. \newline
(\textbf{b}) the topological vertex $\mathcal{T}_{\widehat{\lambda }\widehat{%
\mu }\widehat{\nu }}\left( q,t\right) $ describing the refined version
shifted topological vertex $\mathcal{S}_{\widehat{\lambda }\widehat{\mu }%
\widehat{\nu }}\left( q\right) $. Putting $\widehat{\lambda }=\widehat{\mu }=%
\widehat{\nu }=\emptyset $, we get%
\begin{equation}
\begin{tabular}{|l|}
\hline
$\qquad \mathcal{T}_{3}\left( q,t\right) =\prod\limits_{j=1}^{\infty
}\prod\limits_{i=1}^{\infty }\left( \frac{1+q^{j-1}t^{i}}{1-q^{j-1}t^{i}}%
\right) \qquad $ \\ \hline
\end{tabular}%
\end{equation}%
describing the refined version of Foda-Wheeler relation recovered by setting 
$t=q$.\newline
In the end, notice that it would be interesting to seek whether $\mathcal{T}%
_{\widehat{\lambda }\widehat{\mu }\widehat{\nu }}\left( q,t\right) $ could
be associated with some gauge theory instantons as does $\mathcal{R}%
_{\lambda \mu \nu }\left( q,t\right) $ with the Nekrasov's ones.

\begin{acknowledgement}
: {\small This research work is supported by Protars III D12/25.}
\end{acknowledgement}

\section{Appendix}

In this appendix, we give some useful tools on the \emph{strict} 2d-
partitiions, the \emph{shifted} plane partitions and on Schur functions.%
\newline

\emph{Strict 2d- and shifted 3d- partition}\newline
A 2d- partition, or a Young diagram, denoted as $\lambda =\left( \lambda
_{1},\lambda _{2},\cdots ,\lambda _{r},\cdots \right) $ is a sequence of
decreasing non negative integers $\lambda _{1}\geq \lambda _{2}\cdots \geq
\lambda _{r}\geq \cdots $.\newline
A \emph{strict} 2d- partition\textrm{\ }is a sequence of strictly decreasing
integers\textrm{\ }$\lambda _{1}>\lambda _{2}>\cdots $. The sum of the parts 
$\lambda _{i}$ of the 2d- partition is the weight of $\lambda $ denoted by 
\begin{equation}
\left\vert \lambda \right\vert =\lambda _{1}+\lambda _{2}+\cdots +\lambda
_{r}+\cdots
\end{equation}%
A 2d\ strict partition $\lambda $ is said a partition of n if $\left\vert
\lambda \right\vert =n$ and is represented by its\emph{\ shifted} Young
diagram obtained from the usual diagram by shifting to the right the $i^{th}$
row by $(i-1)$ squares as shown on figure 1.

\begin{figure}[tbph]
\begin{center}
\hspace{-1cm} \includegraphics[width=4cm]{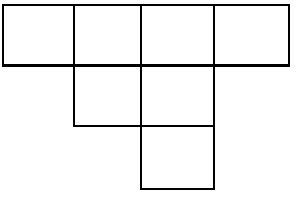}
\end{center}
\par
\vspace{-1cm}
\caption{{\protect\small Shifted Young diagram of }$\protect\lambda =(4,2,1)$%
}
\end{figure}
\ \newline
The \emph{shifted} Young diagram is given by a collection of boxes with
coordinates 
\begin{equation}
\left\{ \left( i,j\right) \mid i=1,...,l(\lambda ),\text{ }i\leq j\leq
\lambda _{i}+i-1\right\} .
\end{equation}%
A \emph{shifted} plane partition $\Pi $ of shape $\lambda $ is determined by
the sequence $(...,.\pi _{-1},\pi _{0},\pi _{1},...)$, where $\pi _{0}$ the
2d- partition on the main diagonal and $\pi _{k}$ is the 2d- partition on
the diagonal shifted by an integer $k$. All diagonal partitions are \emph{%
strict} 2d- partitions forming altogether a \emph{shifted} plane partition
with the property 
\begin{equation}
...\subset \pi _{-1}\subset \pi _{0}\supset \pi _{1}\supset ...
\end{equation}%
For illustration, see the example 
\begin{equation}
\begin{tabular}{llllll}
$\pi _{-2}=\left( 3\right) $ & , & $\pi _{-1}=\left( 4,3\right) $ & , & $\pi
_{0}=\left( 5,3\right) $ & , \\ 
$\pi _{1}=\left( 3,2\right) $ & , & $\pi _{2}=\left( 2\right) $ & , & $\pi
_{3}=\left( 1\right) $ & .%
\end{tabular}
\notag
\end{equation}%
\begin{figure}[tbph]
\begin{center}
\hspace{-1cm} \includegraphics[width=6cm]{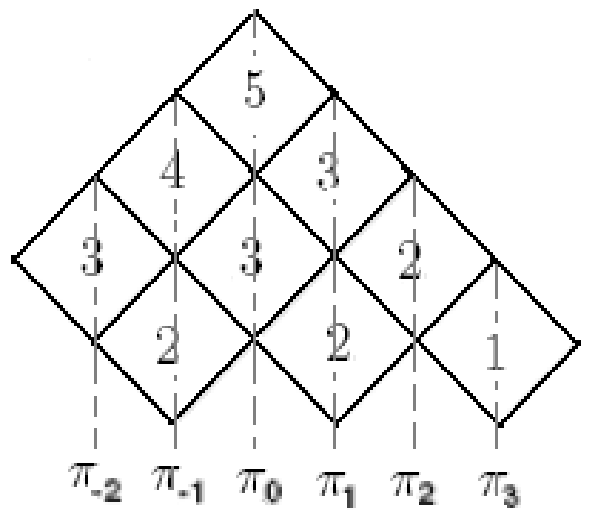}
\end{center}
\par
\vspace{-1cm}
\caption{{\protect\small A strict plane partition }}
\end{figure}

\emph{Property of Schur function for strict partition}\newline
The shifted topological vertex is defined by using skew Schur $P$ and $Q$
functions \textrm{\cite{SBKP}, \cite{shifted}}. These are symmetric
functions that appear in topological amplitudes and are defined by a
sequence of polynomials $P_{\lambda }(x_{1},x_{2},...x_{n})$, $n\in N$, with
the property 
\begin{equation}
P_{\lambda /\mu }(x_{1},\cdots x_{n})=\left\{ 
\begin{array}{c}
\sum\limits_{T}x^{T}\qquad \lambda \supset \mu \\ 
0\qquad \text{otherwise}%
\end{array}%
\right.
\end{equation}%
where the sum is over all shifted Young tableaux of shape $\lambda /\mu $.
The skew Schur function $Q_{\lambda /\mu }$ is related to $P_{\lambda /\mu }$
as in eqs(\ref{38}-\ref{39}). We also have 
\begin{equation}
\sum_{\text{ strict }\lambda }Q_{\lambda }(x)P_{\lambda
}(y)=\prod\limits_{i,j}\left( \frac{1+x_{i}y_{j}}{1-x_{i}y_{j}}\right) .
\end{equation}%
The relation between the Schur function $P_{\lambda }$ for \emph{strict }%
partition $\lambda $ that we have used hereabove and the usual Schur
functions $S_{\tilde{\lambda}}$ for the double partition $\tilde{\lambda}$
is given by 
\begin{equation}
S_{\tilde{\lambda}}(t)=2^{-l(\lambda )}P_{\lambda }^{2}(\frac{t}{2})\text{ ,}
\end{equation}%
where $\frac{t}{2}$ is $(\frac{t_{1}}{2},\frac{t_{3}}{2},\frac{t_{5}}{2}%
,\cdots )$ and $P_{\lambda }(\frac{t}{2})=P_{\lambda }(\frac{t_{1}}{2},\frac{%
t_{3}}{2},\frac{t_{5}}{2},\cdots )$. Notice that the double partition $%
\tilde{\lambda}$ in Frobenuis\textrm{\ }notation reads in terms of the
strict partition $\lambda =(n_{1},n_{2},\cdots ,n_{k})$ as: 
\begin{equation}
\tilde{\lambda}=(n_{1},n_{2},\cdots ,n_{k}\mid n_{1}-1,n_{2}-1,\cdots
,n_{k-1}-1)\text{ .}
\end{equation}

\end{document}